\documentclass[ amsmath, amssymb, aps, 12pt,superscriptaddress]{revtex4-1}

\usepackage{graphicx}
\usepackage{dcolumn}
\usepackage{bm}
\usepackage{hyperref}
\usepackage{xcolor, soul}
\usepackage{braket}

\begin{document}

\title{Measurements of spin-coherence in NV centers for diamond-based quantum sensors}

\author{Lucas Nunes Sales de Andrade}
\affiliation{ Instituto de Física de São Carlos, Universidade de São Paulo (IFSC-USP),
\\ Caixa Postal 369, CEP 13560-970, São Carlos, SP, Brazil. \\ ( srmuniz@ifsc.usp.br -- https://orcid.org/0000-0002-8753-4659 ) }

\author{Charlie Oncebay Segura}
\affiliation{ Instituto de Física de São Carlos, Universidade de São Paulo (IFSC-USP),
\\ Caixa Postal 369, CEP 13560-970, São Carlos, SP, Brazil. \\ ( srmuniz@ifsc.usp.br -- https://orcid.org/0000-0002-8753-4659 ) }
\affiliation{Facultad de Ciencias, Universidad Nacional de Ingeniería, Lima, Peru.}

\author{Sérgio Ricardo Muniz}
\affiliation{ Instituto de Física de São Carlos, Universidade de São Paulo (IFSC-USP),
\\ Caixa Postal 369, CEP 13560-970, São Carlos, SP, Brazil. \\ ( srmuniz@ifsc.usp.br -- https://orcid.org/0000-0002-8753-4659 ) }

\date{31-May-2021}

\begin{abstract}
\vspace{7mm}
One of the biggest challenges to implement quantum protocols and quantum information processing (QIP) is achieving long coherence times, usually requiring systems at ultra-low temperatures. The nitrogen-vacancy (NV) center in diamond is a promising alternative to this problem. Due to its spin properties, easy manipulation, and the possibility of doing optical state initialization and readout, it quickly became one of the best solid-state spin systems for QIP at room temperature. Here*, we present the characterization of the spin-coherence of an ensemble of NV centers in an engineered sample of ultrapure diamond as a testbed for quantum protocols for quantum metrology. 
\\(*) \emph{Paper presented at the Conference \href{https://doi.org/10.1109/SBFotonIOPC50774.2021.9461941}{SBFoton-IOPC-2021}.}
\end{abstract}
\maketitle

\section{Introduction}
In recent years, ensembles of NV centers in diamond became an attractive solid-state platform for quantum technologies. They have been used for nanoscopic mapping of magnetic fields and magnetic resonance microscopy \cite{nanoscale,ultrasensitive,Tetienne2017,thesecharlie}, quantum metrology \cite{observation,quantum}, quantum process tomography, and quantum cryptography \cite{single,experimental}.
    
Implementing quantum protocols and QIP requires long coherence times. Therefore, it is vital to characterize the coherence properties of a given system, determining its characteristic times $\text{T}_1$, $\text{T}_2$ and $\text{T}_2^*$.
Here, we present optical methods and measurements to determine these coherence times for an engineered sample of ultrapure diamond.

\section{Methods}

\subsection{Optical pumping and state initialization.} \label{Section:A}
At room temperature, the thermal equilibrium population of the triplet ground states $^3A_2$, corresponding to the spin states $\ket{0}$ and $\ket{\pm 1}$, are approximately equal, and the ensemble is not spin-polarized. However, one of the remarkable features of NV centers in diamond is how simple it is to produce a long-lived pure spin-polarized state at room temperature, using light. 

\begin{figure}[tb]
    \centering
    \includegraphics[scale=0.85]{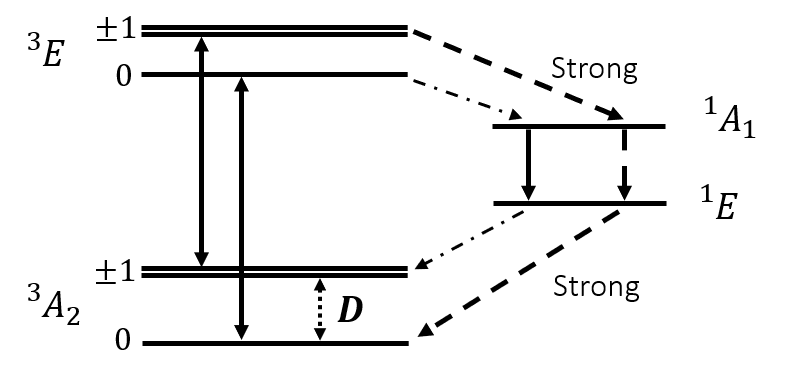}
    \caption{Schematic of the relevant transitions. Solid arrows represent radiative paths, dashed (- -) arrows represent strong non-radiative decays, point-dashed (-.-) arrows indicate weak non-radiative decays, and dotted arrow (...) indicates microwave transition.}
    \label{nvprocess}
\end{figure}

The selection rules for electric dipole transitions are spin conserving but after the optical excitation to $^3E$ (using off-resonance green light at $532\,\textrm{nm}$) the decay involves non-radiative processes that are strongly spin-dependent.
For the excited state $\ket{0}$, the decay happens mostly through a radiative $637\,\textrm{nm}$ band emission \cite{thesepreez}, but for the states $\ket{\pm 1}$ there is a greater probability for intersystem-crossing (ISC) to occur \cite{doherty2013nitrogen}. The intermediate metastable states $^1A_1$ and $^1E$ have lifetimes ranging from $462 \,\ \textrm{to} \,\ 142 \,\ \textrm{ns}$ \cite{acosta2009diamonds,acosta2010optical}, which is at least 10 times larger than the direct radiative decay to the ground state (lifetime $12.9 \pm 0.1 \,\textrm{ns}$) \cite{doherty2013nitrogen}. 
In addition, the decay from the intermediate state $^1E$ is also preferentially to the ground state $\ket{0}$, leading to an effective optical pumping to the ground state $\ket{0}$ after many cycles. 
Therefore, by just shining green light is possible to create a pure state in this system. Figure (\ref{nvprocess}) illustrates these processes. 

\subsection{Microwave interaction.}
The effective Hamiltonian of the ground state for the NV center 
in the $S_z$ basis can be simplified  to \cite{doherty2013nitrogen} 
    
\begin{equation}
    H_{gs} = DS_z^2 + \gamma_e B_z S_z, 
    \label{hamilt}
\end{equation}
where $D = 2.87 \,\ \textrm{GHz}$ is the Zero Field Splitting (ZFS) constant,  
$\gamma_e = 28 \,\ \textrm{MHz/mT}$ is the electron gyromagnetic ratio, and $B_z$ is the magnetic field projection on the NV axis. Since in our system all terms in \eqref{hamilt} are larger than $200\,\textrm{MHz}$, we neglected here terms of order $1\,\ \textrm{MHz}$ or lower, to keep the description simple, for now.
    
The microwave (MW) interaction can be represented by adding a new term to the Hamiltonian \eqref{hamilt}. In the rotating frame of reference, which rotates around the $z$-axis with angular frequency $\omega_{mw}$, the new Hamiltonian is given by

\begin{equation}
    \Tilde{H} = DS_z^2 + (\gamma_e B_z + \omega_{mw})S_z + \frac{\Omega_R}{2\sqrt{2}} S_x,
\end{equation}
where $\Omega_R$ is the Rabi frequency due to the microwave. 

The magnetic (electron-spin) resonance frequency is determined by $\omega_{mw} = D \pm \gamma_e B_z S_z $, leading to transitions $\ket{0} \leftrightarrow \ket{\pm 1}$. Judiciously choosing the values of $B_z$ and $\omega_{mw}$ one can create a qubit, selecting two levels, which here we call states $\ket{0}$ and $\ket{1}$.

The photoluminescence signal, $\mathcal{S}$, is the projection of a general state $\ket{\psi}$ onto the state $\ket{0}$, and results in a signal
\begin{equation}
    \mathcal{S}(t) \propto e^{-\left(t/\text{T}_2^*\right)} \sin^2 \left(\frac{\Omega'_R t}{2}\right),
    \label{rabios}
\end{equation}
where $\Omega'_R$ is the generalized Rabi frequency, $\Omega'_R = \sqrt{\Omega_R^2 + \Delta^2}$, and $\Delta$ is the detuning between the angular frequency $\omega_0 = D \pm \gamma_e B_z$ and the MW frequency $\omega_{mw}$. 
The parameter $\text{T}_2^*$ represents the characteristic decay time of the spin coherence in the presence of local magnetic inhomogeneities.
    
\subsection{Spin-coherence times and spin dynamics.}    \label{Section:C}
To measure the spin-coherence times, we use well-established sequences from NMR. Here, in addition to the usual pulse sequences to observe Rabi oscillations, we use two particular sequences to measure the characteristic times $\text{T}_1$ and $\text{T}_2$, corresponding respectively to the longitudinal and transversal coherence times. 

\subsubsection{Longitudinal relaxation, ($\text{T}_1$)}    
In thermal equilibrium, the populations of the states $\ket{0}\,\text{and}\,\ket{\pm 1}$ are given by the corresponding Boltzmann factor, almost equally populated for the ground state at room temperature. After the optical pumping, described in section \ref{Section:A}, the system is spin-polarized into the $\ket{0}$ state. Applying a $\pi$-pulse will invert the population to the state $\ket{1}$, but after some time, the system returns to the thermal equilibrium state. 
This relaxation process is represented by the time constant $\text{T}_1$, called longitudinal relaxation or spin-lattice relaxation constant.

To measure $\text{T}_1$ one simply applies a $\pi$-pulse and waits for some time delay $\tau$ to perform a measurement ($S_z$) of the spin projection in the z-direction. In our system, it corresponds to a simple one-pulse sequence ($\pi - \tau $), and the photoluminescence signal decays in time as
\begin{equation}
    \mathcal{S}(t) \propto e^{\left(-t/\text{T}_1\right)}.
       \label{t1eq}
\end{equation}

\subsubsection{Transversal relaxation, ($\text{T}_2$, $\text{T}_2^*$)}
It can be separated into two types: the homogeneous dephasing characterized by the constant $\text{T}_2$, and the inhomogeneous dephasing characterized by $\text{T}_2^*$. Many mechanisms contribute to transversal relaxation, for example, the interaction between individual spins (electronic NV spin and the nuclear spin from nitrogen and nearby atoms, usually $^{13}C$); temperature fluctuations; strain gradients and inhomogeneities on the external magnetic field. These mechanisms cause incoherent dephasing, a process called spin-spin relaxation or transversal relaxation.

One way to measure $\text{T}_2$, removing the inhomogeneous dephasing, is to use a special pulse sequence called Hahn echo, or spin-echo.
In our case, the Hahn echo sequence has three MW pulses ($\pi/2 - \tau - \pi - \tau - \pi/2$).
The first $\pi/2$-pulse creates a superposition of the states, geometrically interpreted as a state vector in the transverse $xy$-plane on the Bloch sphere. The state is allowed to evolve freely for a time $\tau$. During that time, the phase at each point evolves according to its local field, and it may cause dephasing between different points. Applying a $\pi$-pulse allows to revert the accumulated phase after waiting for another time interval $\tau$, effectively refocusing (i.e., synchronizing) the spins onto the $x$-axis. A final $\pi/2$-pulse rotates the state vector onto the $z$-axis for optical readout.

Varying the time interval $\tau$, we can determine $\text{T}_2$ by fitting the data with the following equation \cite{thesecharlie}:
\begin{equation}
\begin{split}
    \mathcal{S}(\tau) & = e^{-\left(2\tau/\text{T}_2\right)^n} [1 - 0.25k(2-2\cos (\omega_a \tau) - 2 \cos (\omega_b \tau)  \\
        & + \cos (\omega_a + \omega_b)\tau) + \cos (\omega_a - \omega_b)\tau))],
        \label{t2eq}
\end{split}
\end{equation}
where $\omega_a, \omega_b$ are the angular frequencies involving the nuclear spin, $k$ depends on the external magnetic field $\mathbf{B}$, $\text{T}_2$ is the coherence time and $n$ is a power index that varies depending on the decoherence mechanism \cite{thesecharlie}.
    
The hierarchy for the time scales is given by
\begin{equation}
    \text{T}_1 \geq \text{T}_2 > \text{T}_2^*,
\end{equation}
where the coherence time $\text{T}_2^*$ is always smaller than $\text{T}_2$ and $\text{T}_2$ can be less than or equal to $\text{T}_1$.

\section{Experimental setup}
\subsection{Microwave generation.}
The microwave is produced by a frequency synthesizer (Stanford SG384), connected to a fast switch (CMCCS0947A-C2), that receives a TTL signal from a fast digital card (SpinCore, PulseBlaster, PBESR-PRO-300), with a time resolution of $3.3 \,\ \textrm{ns}$. The signal generated by the synthesizer is amplified by a $45\,\textrm{dB}$ power amplifier (Mini-Circuits ZHL-16W-43-S+). A circulator (Circulator CS 3000), with a $50 \,\Omega$ heavy-duty resistive load, is used to protect the system.

\subsection{Optical Setup.}    
The experiment can operate in two modes: continuous and pulsed modes. We use an acousto-optic modulator (AOM) (Crystal Technology, AOMO 3110-320) to control and switch between both modes. When the AOM is (de)activated by a radio frequency (RF) signal, it turns on(off) the green laser (532 nm) light, used to excite the NV centers. The AOM alignment was optimized to guarantee the fastest on/off switching of the light.

To control the AOM activation, we built an electronic circuit that changes the RF's frequency and amplitude. Choosing the right parameters, we maximize the first-order diffraction to use it as the excitation beam. The circuit is connected to a fast RF switch (Mini-circuits, ZASWA-2-50DR+) controlled by the SpinCore card (PBESR-PRO-300). The RF for the AOM is amplified using a Mini-circuits (ZHL-03-5WF) amplifier, and directed to the AOM. By selecting only the first order deflection with an iris, we use this setup to turn the laser on and off.
    
For the optical excitation, we use a green laser at $532\,\textrm{nm}$ (Thorlabs, DJ532-40). The beam position on the AOM was adjusted to achieve the minimum delay ($\approx130\,\textrm{ns}$) and the fastest response time ($\approx35\,\textrm{ns}$), typically with a total switching time of  $165\,\textrm{ns}$. Since this time delay is fixed and stable, we simply adjust the pulse delays to account for it in the time sequence, obtaining reliable results. 
After the AOM and the iris, a pair of lenses of $25 \,\textrm{mm}$ and $5\,\textrm{mm}$ focal lengths is used to expand and collimate the beam. A dichroic (SemRock, Di02-R561-25x36) separates the excitation beam and the photoluminescence emitted by the diamond. 

\begin{figure}[t]
\centering
\includegraphics[scale = 0.5]{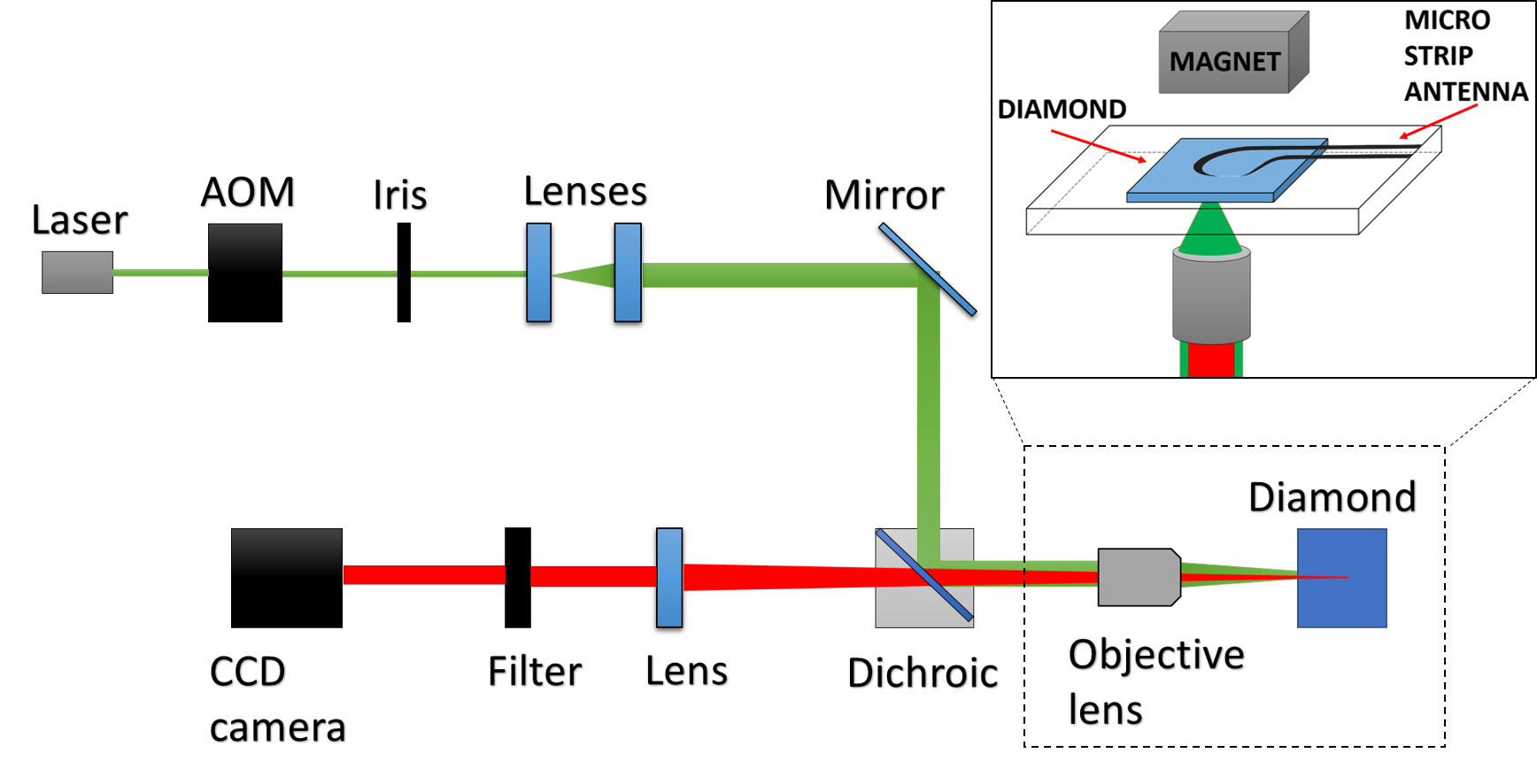}
\caption{A $532\,\textrm{nm}$ laser beam goes through an AOM, and an iris selects the first diffracted order to go through a telescope. The dichroic mirror reflects green light towards the objective, focusing it onto the sample and collecting the fluorescence that goes through the dichroic and a filter before being captured by a CCD camera. The boxed (dashed) region illustrates the details (not scaled) near the sample.}
\label{esquema}
\end{figure}

The dichroic reflects light with wavelengths shorter than $561 \,\textrm{nm}$, that goes through the objective lens (Apochromat 50x, NA=0.95), which focuses the light onto the diamond that fluoresces. The same objective captures and collimates the fluorescence that goes through the dichroic and a final lens, forming an image onto a CCD camera (Point Grey, FL3-FW-0S1M-C). An optical filter (ELH0550) was positioned right before the camera to block any green light. Figure \ref{esquema} shows a schematic of the setup.

\subsection{Optically Detected Magnetic Resonance.}
The Optically Detected Magnetic Resonance (ODMR) is the primary technique used in the experiments. Due to the spin dynamics described in section \ref{Section:A}, microwave at a resonance frequency ($\omega_0 = D \pm \gamma_e B_z$) drives the transition $\ket{0} \rightarrow \ket{1}$, causing a decrease in the fluorescence signal. We detect this signal change using a CCD camera.

Typically, the first step in the experimental sequence is to obtain ODMR spectra, scanning the microwave frequency to find the resonances. In our setup, a permanent neodymium magnet near the diamond provides a bias field $B_z=85\,\textrm{G}$ to lift the degeneracy of the states $\ket{\pm 1}$. We can choose anyone out of 8 different resonances due to 4 different orientations of the NV axis. For the experiments described here, we used a resonance peak at $3111\,\textrm{MHz}$.

\subsection{Pulse sequences and data collection.}

The experiments follow the methods described in section \ref{Section:C}.
A  $350\,\mu\textrm{s}$ optical pumping initializes the sequences at the state $\ket{0}$. Then, one of the pulse sequences is applied, controlled by digital pulses from the SpinCore card. Right after the MW sequence, a light pulse provides readout and optical pumping to restart a new sequence.

Using a Rabi sequence, we determine the $\pi$-pulse (i.e., the time to invert the population to $\ket{1}$), indicated by the minimum fluorescence. This experiment also gives us information about the $\text{T}_2^*$, by fitting the data with \eqref{rabios}.
    
To measure $\text{T}_1$, after the initialization, we apply a $(\pi - \tau)$ sequence and determine $\text{T}_1$ from fitting the signal $\mathcal{S}(t)$ using \eqref{t1eq}, as shown in Fig. \ref{graft1}.
    
For $\text{T}_2$ we use the Hahn-echo sequence, given in section \ref{Section:C}, and used \eqref{t2eq} to determine $\text{T}_2$, as shown in Fig. \ref{graft2l}.
    
To collect and process the data, we capture three different images for each value of $\tau$. The first image collects the fluorescence using the complete sequence, including microwave and light pulses. The second image has an identical time sequence but does not turn on the microwave (i.e., it has only the laser pulses), and it is used for normalization. The last image contains only the background reading from the CCD, using the same time sequence, without laser or MW pulses. The third image is subtracted (point-to-point) from the other two images before the normalized image is calculated.

The measurement of these three images is collected one after the other, constituting one measurement block. This is done to minimize the effects of eventual (slow) drifting of the experimental parameters. Each measurement block is repeated 150 times at each value of $\tau$, and averaged to reduce the measurement noise. Therefore, each data point shown in the figures is the averaged value of 150 measurement blocks.

\section{Results and Discussions.}
\begin{figure}[t]
\centering
\includegraphics[scale=0.49]{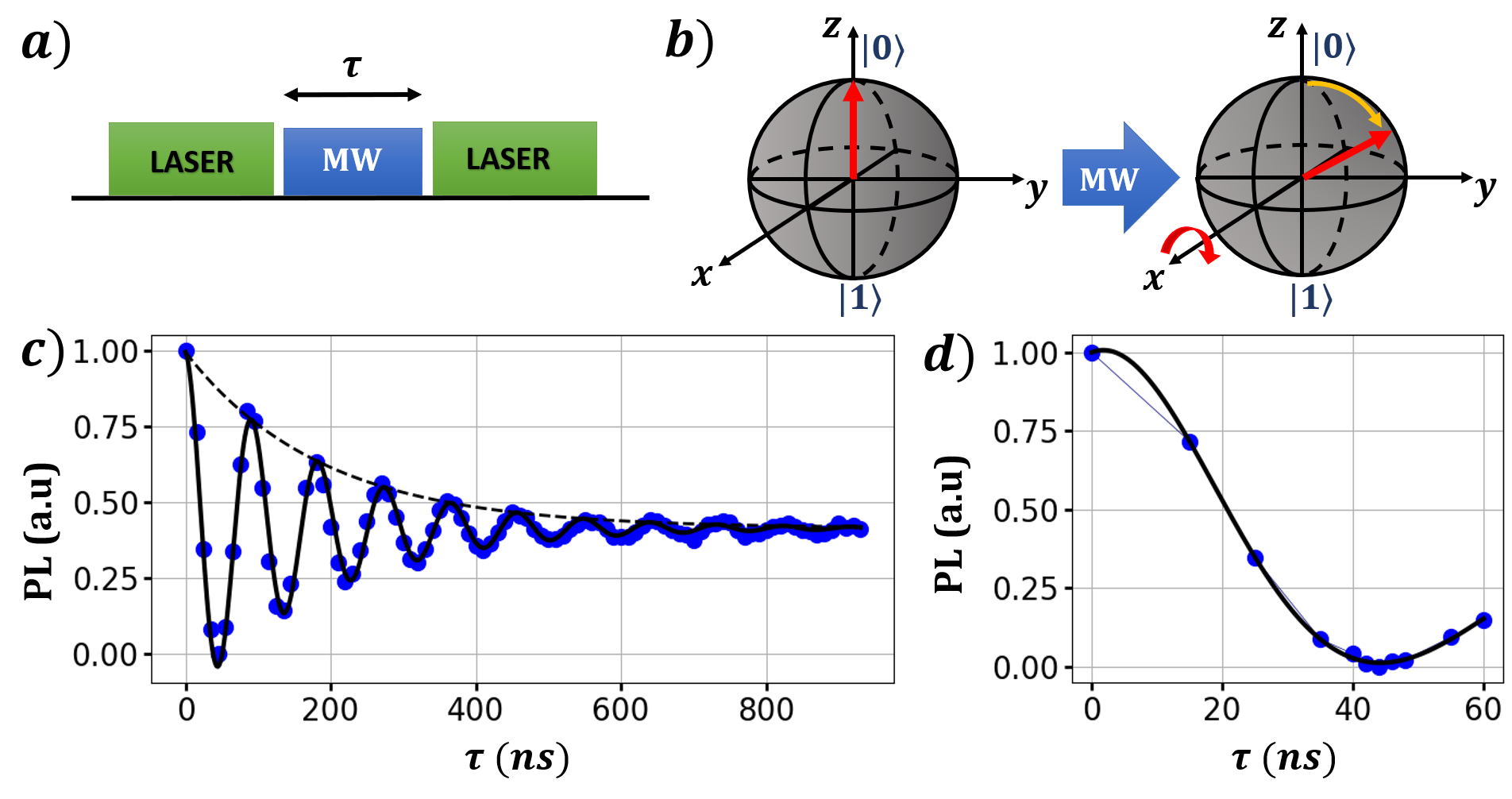}
\caption{a) Pulse sequence to measure Rabi oscillation. b) Representation on Bloch sphere. c) Experimental (blue) and fitted (black) data. d) A closer look at the minimum photoluminescence to determine the $\pi$-pulse.}
\label{graftrabi1}
\end{figure}

The first pulse sequence implemented was the Rabi coherent oscillation of the qubit produced in electron spin of each NV center of the ensemble, in the region excited by the laser. Figure \ref{graftrabi1} shows the result using a MW power of $40\,\textrm{dBm}$, highlighting the measurements near the time corresponding to a $\pi$-pulse of $t_{\pi} = 44\,\textrm{ns}$.

\begin{figure}[tb]
    \centering
    \includegraphics[scale=0.67]{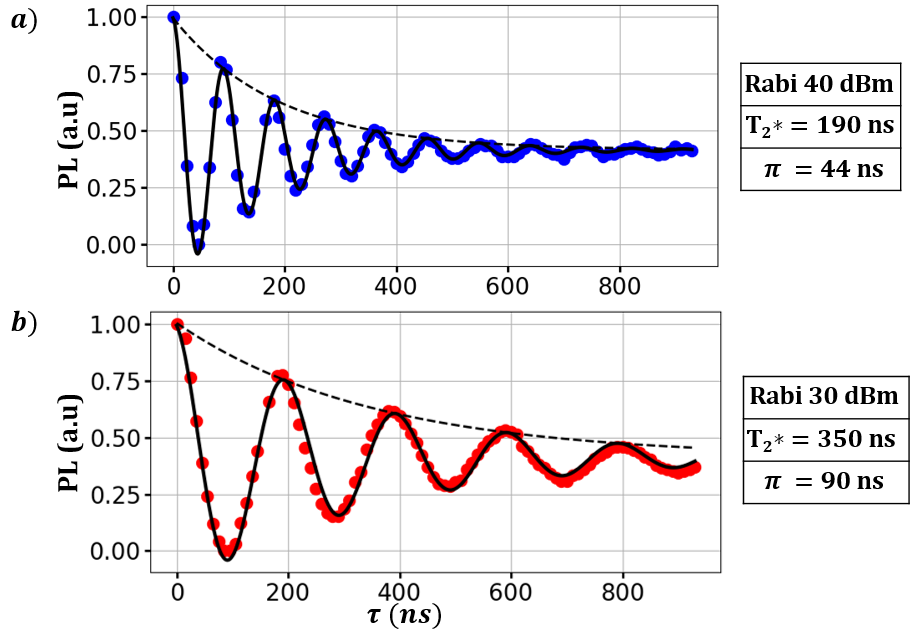}
    \caption{Variation of $\pi$ pulse and $\text{T}_2^*$ with the power. The dependence with the MW power indicates larger inhomogeneities, possibly due to the larger spectral width of the pulse exciting larger areas in the sample. }
    \label{graftrabi2}
\end{figure}

As expected, the greater the MW power, the shorter is the $\pi$-pulse duration, because the Rabi frequency is proportional to the intensity of the magnetic field $B_{mw}$. Interestingly, the coherence time $\text{T}_2^*$ also changes with power, as shown in Fig. \ref{graftrabi2}. The values vary from $\text{T}_2^* = 350 \,\ \textrm{ns}$ at $30 \,\ \textrm{dBm}$ to $\text{T}_2^* = 190 \,\ \textrm{ns}$ at $40 \,\ \textrm{dBm}$. This is behavior may be due to the broader spectral composition of the MW pulse, leading to interactions with a larger region of spins in the sample and possibly larger inhomogeneities, including the MW field.

\begin{figure}[tb]
    \centering
    \includegraphics[scale=0.75]{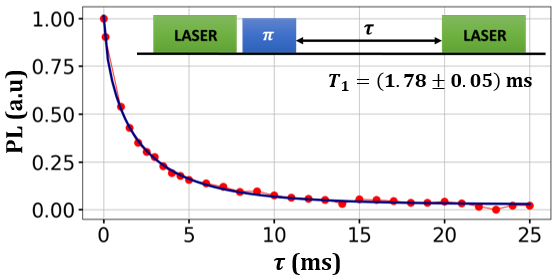}
    \caption{Spin-Lattice relaxation measurement. The pulse sequence can be seen in the top right corner. Data in red and fit in blue.}
    \label{graft1}
\end{figure}

Determined the $\pi$-pulse, we use it in the $\text{T}_1$ sequence. The time delay was varied from $0$ to $25\,\textrm{ms}$, observing the signal decaying exponentially, as seen in Fig. (\ref{graft1}). Fitting the data with \eqref{t1eq} we calculate $\text{T}_1 = 1.78 \pm 0.05 \,\ \textrm{ms}$. This value does not depend on the MW power and is consistent with the literature \cite{doherty2013nitrogen} for our sample quality of (type IIa. diamond) at room temperature. See references \cite{thesecharlie,charlie2021} for details on the NV-sensor (sample) preparation. 

\begin{figure}[tb]
    \centering
    \includegraphics[scale=0.7]{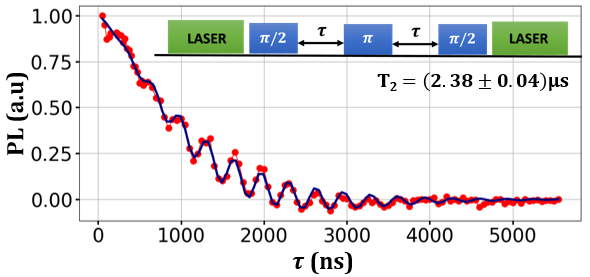}
    \caption{Spin-Spin Relaxation measurement. The microwave sequence is shown in the top right corner. Experimental data in red (points) and fit of the modulated decay in blue. One can notice the presence of oscillations due to hyperfine interaction between electron and nuclear spins.}
    \label{graft2l}
\end{figure}

The result of the Hahn Echo sequence is shown in Fig. \ref{graft2l}, displaying a clear modulation on top of the decay curve. This modulation is due to the hyperfine interaction, as studied in \cite{thesecharlie}. 
The calculated transversal relaxation is $\text{T}_2 = 2.38 \pm 0.04\,\mu\textrm{s}$ and is not as sensitive to the MW power as $\text{T}_2^*$. Fitting using (\ref{t2eq}), we obtain  $f_a = \omega_a/2\pi = (3.04 \pm 0.01)\,\textrm{MHz}$, $f_b = \omega_b/2\pi =  (70 \pm 30)\,\textrm{kHz}$, $n = 1.29 \pm 0.04$ and $k = 3$. The value of $f_a$ corresponds to the hyperfine parameter $A^{\parallel} \approx 3.0 \,\ \textrm{Mhz}$, and $f_b$ is consistent with the nuclear Larmor frequency.

\vspace{-1mm}

\section{Conclusions.}
Here, we described how to measure the spin-coherence of an ensemble of NV centers optically and presented results for our system: 
$\text{T}_1 = 1.78 \pm 0.05 \,\ \textrm{ms}$, $\text{T}_2 = 2.38 \pm 0.04 \,\ \mu\textrm{s}$ and $\text{T}_2^* = 350 \,\ \textrm{ns}$ at $30 \,\ \textrm{dBm}$. These results are consistent with the literature for the parameters of our engineered sample \cite{thesecharlie, charlie2021} and provide good coherence times for a quantum system operating at room temperature. 

The presence of hyperfine interaction provides a second qubit at our disposal. In addition to the electron spin of the NV center, we also have access to the nuclear spin of the nearby $^{15}\textrm{N}$ nucleus, creating the possibility for 2-qubit operations.
This allows us to start exploring QIP and applying quantum protocols for metrology and fundamental studies currently underway in our laboratory.

\section*{Acknowledgment}
We thank Victor Acosta for providing the diamond sample used in this study and $\,\,$Eduardo de Azevêdo for helpful discussions. The authors acknowledge financial support from CAPES (process 88887.372074/2019-00), CNPq (process 141453/2021-4), and FAPESP (grants 2019/27471-0 and 2013/07276-1).


\begin{thebibliography}{13}

\bibitem{nanoscale}
Gopalakrishnan Balasubramanian, I.~Y. Chan, Roman Kolesov, Mohannad Al-Hmoud,
  Julia Tisler, Chang Shin, Changdong Kim, Aleksander Wojcik, Philip~R. Hemmer,
  Anke Krueger, Tobias Hanke, Alfred Leitenstorfer, Rudolf Bratschitsch, Fedor
  Jelezko, and J{\"{o}}rg Wrachtrup.
\newblock Nanoscale imaging magnetometry with diamond spins under ambient
  conditions.
\newblock {\em Nature}, 455(7213):648, 2008.

\bibitem{ultrasensitive}
Liam~T Hall, Charles~D Hill, Jared~H Cole, and Lloyd~CL Hollenberg.
\newblock Ultrasensitive diamond magnetometry using optimal dynamic decoupling.
\newblock {\em Physical Review B}, 82(4):045208, 2010.

\bibitem{Tetienne2017}
Jean-Philippe Tetienne, Nikolai Dontschuk, David~A. Broadway, Alastair Stacey,
  David~A. Simpson, and Lloyd C.~L. Hollenberg.
\newblock Quantum imaging of current flow in graphene.
\newblock {\em Science Advances}, 3(4), 2017.

\bibitem{thesecharlie}
Charlie~Oncebay Segura.
\newblock {\em Diamond studies for applications in quantum technologies}.
\newblock PhD thesis, Universidade de São Paulo, Instituto de Física de São
  Carlos, 2019.
\\ \href{https://doi.org/10.11606/T.76.2019.tde-01082019-152208}{https://doi.org/10.11606/T.76.2019.tde-01082019-152208}

\bibitem{observation}
Fedor Jelezko, T~Gaebel, I~Popa, M~Domhan, A~Gruber, and Jorg Wrachtrup.
\newblock Observation of coherent oscillation of a single nuclear spin and
  realization of a two-qubit conditional quantum gate.
\newblock {\em Physical Review Letters}, 93(13):130501, 2004.

\bibitem{quantum}
MV~Gurudev Dutt, L~Childress, L~Jiang, E~Togan, J~Maze, F~Jelezko, AS~Zibrov,
  PR~Hemmer, and MD~Lukin.
\newblock Quantum register based on individual electronic and nuclear spin
  qubits in diamond.
\newblock {\em Science}, 316(5829):1312--1316, 2007.

\bibitem{single}
Alexios Beveratos, Rosa Brouri, Thierry Gacoin, Andr{\'e} Villing,
  Jean-Philippe Poizat, and Philippe Grangier.
\newblock Single photon quantum cryptography.
\newblock {\em Physical review letters}, 89(18):187901, 2002.

\bibitem{experimental}
Romain All{\'e}aume, Fran{\c{c}}ois Treussart, Ga{\"e}tan Messin, Yannick
  Dumeige, Jean-Fran{\c{c}}ois Roch, Alexios Beveratos, Rosa Brouri-Tualle,
  Jean-Philippe Poizat, and Philippe Grangier.
\newblock Experimental open-air quantum key distribution with a single-photon
  source.
\newblock {\em New Journal of physics}, 6(1):92, 2004.

\bibitem{thesepreez}
L.~Du Preez.
\newblock {\em Electron paramagnetic resonance and optical investigations of
  defect centre in diamond}.
\newblock PhD thesis, University of Witwaters, 1965.

\bibitem{doherty2013nitrogen}
Marcus~W Doherty, Neil~B Manson, Paul Delaney, Fedor Jelezko, J{\"o}rg
  Wrachtrup, and Lloyd~CL Hollenberg.
\newblock The nitrogen-vacancy colour centre in diamond.
\newblock {\em Physics Reports}, 528(1):1--45, 2013.

\bibitem{acosta2009diamonds}
Victor~M Acosta, Erik Bauch, Micah~P Ledbetter, Charles Santori, K-MC Fu,
  Paul~E Barclay, Raymond~G Beausoleil, H{\'e}lo{\"\i}se Linget, Jean~Francois
  Roch, Francois Treussart, et~al.
\newblock Diamonds with a high density of nitrogen-vacancy centers for
  magnetometry applications.
\newblock {\em Physical Review B}, 80(11):115202, 2009.

\bibitem{acosta2010optical}
VM~Acosta, A~Jarmola, E~Bauch, and D~Budker.
\newblock Optical properties of the nitrogen-vacancy singlet levels in diamond.
\newblock {\em Physical Review B}, 82(20):201202, 2010.

\bibitem{charlie2021}
Charlie {Oncebay Segura} and Sergio~R. Muniz.
\newblock Diamond-based optical vector magnetometer.
\newblock In {\em 2021 SBFoton International Optics and Photonics Conference
  (SBFoton IOPC)}, São Carlos, Brazil, May 2021.
\newblock \href{https://doi.org/10.1109/sbfotoniopc50774.2021.9461950}{https://doi.org/10.1109/sbfotoniopc50774.2021.9461950}

\end{thebibliography}
\end{document}